\documentclass[aps,prd,nofootinbib,showkeys,preprint,floatfix]{revtex4}

\usepackage{amssymb}
\usepackage{amsmath}
\usepackage{amsfonts}
\usepackage{graphicx}
\usepackage{color}
\usepackage{xspace}
\usepackage{ulem}
\usepackage{lscape}
\usepackage{ wasysym }

\usepackage{multirow,rotating}

\usepackage[pdfborder={0 0 1}]{hyperref}
\definecolor{redLinks}{rgb}{0.6, 0, 0}
\definecolor{greenLinks}{rgb}{0, 0.6, 0} 
\definecolor{blueLinks}{rgb}{0, 0, 0.6}

\def\gsim{\raise0.3ex\hbox{$\;>$\kern-0.75em\raise-1.1ex\hbox{$\sim\;$}}}
\def\lsim{\raise0.3ex\hbox{$\;<$\kern-0.75em\raise-1.1ex\hbox{$\sim\;$}}}

\def\znbb{0\nu\beta\beta}

\newcommand{\beq}{\begin{equation}}
\newcommand{\eeq}{\end{equation}}
\newcommand{\bea}{\begin{eqnarray}}
\newcommand{\eea}{\end{eqnarray}}
\def\eq#1{{eq.~(\ref{#1})}}
\def\eqs#1#2{{eqs.~(\ref{#1})--(\ref{#2})}}

\newcommand{\ba}[1]{\begin{eqnarray} \label{(#1)}}
\newcommand{\ea}{\end{eqnarray}}

\newcommand{\AddrAHEP}{
  {\it AHEP Group, Instituto de F\'{\i}sica Corpuscular --
    C.S.I.C./Universitat de Val{\`e}ncia \\
    Edificio Institutos de Investigacion, Parc Cientific de Paterna, 
  Apartado 22085,
  E--46071 Val{\`e}ncia, Spain}}

\newcommand{\AddrINFN}{
INFN, Laboratori Nazionali di Frascati \\ 
Via Enrico Fermi 40, 00044 Frascati, Italy\\ }

\def\gsim{\raise0.3ex\hbox{$\;>$\kern-0.75em\raise-1.1ex\hbox{$\sim\;$}}}
\def\lsim{\raise0.3ex\hbox{$\;<$\kern-0.75em\raise-1.1ex\hbox{$\sim\;$}}}

\begin{document}

\preprint{IFIC/16-48}

\title{Quasi-Dirac neutrinos at the LHC}

\author{G. Anamiati} \email{anamiati@ific.uv.es}\affiliation{\AddrAHEP}
\author{M. Hirsch} \email{mahirsch@ific.uv.es}\affiliation{\AddrAHEP}
\author{E. Nardi} \email{enrico.nardi@lnf.infn.it}\affiliation{\AddrINFN}


\begin{abstract}
  Lepton number violation is searched for at the LHC using same-sign
  leptons plus jets. The standard lore is that the ratio of same-sign
  lepton to opposite-sign lepton events, $R_{ll}$, is equal to
  $R_{ll}=1$ ($R_{ll}=0$) for Majorana (Dirac) neutrinos. We clarify
  under which conditions the ratio $R_{ll}$ can assume values
  different from 0 and 1, and we argue that the precise value $0 <
  R_{ll} < 1$ is controlled by the mass splitting versus the width of
  the quasi-Dirac resonances. A measurement of $R_{ll}\neq 0,1$ would
  then contain valuable information about the origin of neutrino
  masses.  We consider as an example the inverse seesaw mechanism in a
  left-right symmetric scenario, which is phenomenologically
  particularly interesting since all the heavy states in the high
  energy completion of the model could be within experimental reach.
  A prediction of this scenario is a correlation between the values of
  $R_{ll}$ and the ratio between the rates for heavy neutrino decays
  into standard model gauge bosons, and into three body final states
  $ljj$ mediated by off-shell $W_R$ exchange.
\end{abstract}

\maketitle

\tableofcontents


\section{Introduction}
\label{sect:Intro}

The tiny values of the standard model (SM) neutrino masses can be more
elegantly explained under the assumption that neutrinos are Majorana
particles.  Majorana neutrinos necessarily imply lepton number
violation (LNV), a well known LNV process is for example neutrinoless
double beta decay (for reviews on $\znbb$ see for example
\cite{Avignone:2007fu,Deppisch:2012nb}).  LNV is also searched for at
the LHC, using as a signature final states containing two same-sign
(SS) leptons (plus jets and no missing energy in the event). This
signature, specific for collider searches, was originally proposed in
\cite{Keung:1983uu} in the context of left-right (LR) symmetric
extensions of the standard model
(SM)~\cite{Pati:1974yy,Mohapatra:1974gc,Mohapatra:1980yp}.\footnote{Although
  it is not widely known, SS dilepton events are not a distinctive
  feature of LR scenarios.  They can also arise, in principle, in a
  variety of LNV models \cite{Helo:2013ika} some of which do not
  introduce right-handed neutrinos.}
 
A heavy Majorana neutrino, once produced on mass-shell, decays with
equal probabilities to either a lepton ($l^{-}$) or an anti-lepton
($l^{+}$) (plus, for example, jets). Therefore, for dilepton events
produced via $W\to l N \to lljj$ a ratio of SS to opposite sign (OS) 
dileptons $R_{ll}=1$ is expected.\footnote{Via loop corrections small
  departures from exact $R_{ll}\equiv 1$ are possible. This signals CP
  violation and is a necessary ingredient for models of leptogenesis
  \cite{Fukugita:1986hr} (see \cite{Buchmuller:2004nz,Davidson:2008bu,
    Fong:2013wr} for reviews).}  For a Dirac neutrino $R_{ll}=0$ since
lepton number is conserved.  In this paper we point out that in models
with so-called ``quasi-Dirac'' neutrinos, $R_{ll}$ can instead assume
any value in the interval [0,1]. Hence a measurement of $R_{ll}$, 
different from zero or one, would provide valuable informations on the
mechanism underlying the generation of neutrino masses.  Let us recall
that ``quasi-Dirac'' refers to a pair of Majorana neutrinos with a
small mass splitting and a relative CP-sign between the two states,
and that would correspond to a Dirac neutrino in the limit of exact
mass degeneracy.  Pairs of quasi-Dirac neutrino often appear in
seesaw-type models at scales not far from the electroweak scale,
such as the inverse \cite{Mohapatra:1986bd} and the linear
\cite{Akhmedov:1995vm,Akhmedov:1995ip} seesaw, so that the possibility
of observing $R_{ll}\neq 1,0$ is naturally interweaved with the
possibility of producing new heavy neutrinos in high energy
collisions.\footnote{Scenarios with quasi degenerate right-handed
  neutrinos with masses and couplings allowing for their production at the
  LHC, but of the Dirac type~\cite{Bray:2007ru}, or effectively
  yielding lepton  number  conservation~\cite{Kersten:2007vk},  have
  been also proposed.}

Both, the ATLAS \cite{ATLAS:2012ak,Aad:2015xaa} and the CMS
collaboration \cite{CMS:2012uwa,Khachatryan:2014dka} have published
results for dilepton plus jets $\ell\ell jj$ events.  In general, the
sensitivities of ATLAS and CMS are quite similar.  Nevertheless, there
are some important differences in the analysis of the two
collaborations.  ATLAS, in its first publication \cite{ATLAS:2012ak},
gave results for both, SS and OS lepton events separately. Since no
excess was observed and the background in the OS sample is
considerably larger than in the SS sample, the limits derived from the
combined data are dominated by the SS sample.  Note that this
combination assumes implicitly $R_{ll}=1$.  Probably for this reason,
in the latest analysis \cite{Aad:2015xaa} ATLAS gives only the limits
derived from the SS sample. CMS, on the other hand, gives only
combined results for OS and SS samples
\cite{CMS:2012uwa,Khachatryan:2014dka}, despite the fact that CMS
routinely measures the lepton charge.  In the latest CMS analysis,
which uses the full $\sqrt{s}=8$ TeV statistics
\cite{Khachatryan:2014dka}, an excess in the electron sample around
$m_{eejj} \simeq 2$ TeV was reported. The excess contains 14 events
with an estimated background of 4 events, corresponding to a local
significance of about $2.8$ $\sigma$ c.l. No excess was observed in
the muon sample. CMS points out that (i) only one of the 14 events is
SS and (ii) no localized excess in $m_{\ell_2jj}$, as would be
expected from the decay of an on-shell intermediate $N$, is observed,
and thus it was concluded that the excess is not consistent with the
expectations from LR symmetric models. ATLAS, on the other hand, has
zero events in the same invariant mass bin, but since in
\cite{Aad:2015xaa} ATLAS does not provide results for OS dileptons,
their result is not inconsistent with CMS.  The CMS excess has caused
a flurry of theoretical activity~\cite{referringCMS}, several of the
proposed explanations are based on LR symmetric models, see for
example the works in~\cite{Deppisch:2014qpa,Heikinheimo:2014tba,%
  Dobrescu:2015qna,Brehmer:2015cia,Cheung:2015nha}, where however
$R_{ll}=1$ is generally expected. Note that $R_{ll}=0$ is expected in
LR models with a linear seesaw \cite{Deppisch:2015cua}, while $R_{ll}
< 1$ can be obtained in the $R$-parity violating supersymmetric model
of~\cite{Allanach:2014lca}.  However, particularly relevant for our
study is \cite{Dev:2015pga} which also focuses on a LR symmetric model
equipped with the inverse seesaw mechanism, and where it is stressed
that heavy pseudo-Dirac neutrinos allow to arrange for a suppression
of SS versus OS dilepton events, and hence for a value of $R_{ll}<1$.
Although we agree on the general statement, we find disagreement as
concerns the dependence of $R_{ll}$ on the relevant model parameters.
In particular, differently from~\cite{Dev:2015pga}, we find that the
value of $R_{ll}$ does not display a parametric dependence on the
overall right-handed (RH) neutrino mass scale.\footnote{The authors of
  \cite{Deppisch:2015qwa} study the inverse seesaw within the standard
  model group. We agree with their expression for the LNV
  amplitude. However, different from the LR case, which we study in
  this paper, \cite{Deppisch:2015qwa} concludes that LNV events are
  not observable for heavy neutrino masses above 100 GeV in their
  setup.}

Neutrino oscillation experiments have established that neutrino flavor
numbers are not conserved. By now we have very precise information on
the active neutrino mixing angles, see for example
\cite{Forero:2014bxa}.  Basically the ``solar'',
$\sin^2\theta_{\odot}\simeq 1/3$, and ``atmospheric'',
$\sin^2\theta_{\rm Atm}\simeq 1/2$, angles are large, while the
``reactor'' angle, $\sin^2\theta_{\rm R}\simeq 0.0234$, is smaller. It
is therefore quite unnatural to assume that heavy neutrinos, if they
exist, would only decay to the same lepton flavor associated with
their production (as for example in
$W^+_R \to \ell_j^+ N_R \to \ell_j^+\ell_j^- W_R^*$).  From the
theoretical point of view, different flavor dilepton events
$\ell^+_i\ell^-_j$ and $\ell^+_i\ell^+_j$ with $i\neq j$ are expected
to contribute sizeably to the whole dilepton samples, and for some
choices of the model parameters they could even dominate the total
signal.  The relative amount of different flavor dilepton events could
also provide valuable information about the structure of the seesaw
matrices.  Unfortunately, both ATLAS and CMS use $e\mu$ dilepton
samples to estimate the backgrounds, giving results only for $ee$ and
$\mu\mu$ samples separately.  We would like to stress that different
flavor dilepton events should also be considered as a possible signal,
and that presenting experimental results separately for each specific
flavor channel would provide additional valuable information.

This paper is organized as follows. In the next section, we recall the
main features of the inverse seesaw model~\cite{Mohapatra:1986bd}, we
describe in some details the steps to achieve approximate
diagonalization of the full $9\times 9$ neutrino mass matrix, and we
write down the heavy neutrino couplings to the LR gauge bosons and to
the Higgs. In the same section we also introduce a convenient
parametrization which, in the inverse seesaw, plays an analogous role
as the Casas-Ibarra parametrization~\cite{Casas:2001sr} in the
type-I seesaw.  In section \ref{sect:Rratio} we derive the expression
for the ratio $R_{ll}$. Our result shows that the condition required
for obtaining values of $R_{ll}\neq 0,1$ is that the mass degeneracy
of the quasi-Dirac neutrino pairs must be of the order of their decay
width. In  section \ref{sect:Pheno} we discuss all relevant phenomenology
(two and three body decays and branching ratios) that could be
measured 
at the LHC.  We close with a short summary.

\section{The inverse seesaw}
\label{sect:ISS}


In this section we discuss the inverse seesaw mechanism. In subsection
\ref{subsect:Setup} we present the inverse seesaw mass matrix and
parameter counting, in~\ref{subsect:Appr} we describe an approximate
diagonalization procedure for the $9\times 9$ mass matrix,
in~\ref{subsect:cpl} we give the neutrino couplings to gauge (and
Higgs) bosons, and in~\ref{subsect:CI} we provide a re-parametrization
of the inverse seesaw that allows to fulfill automatically the
experimental constraints from low-energy neutrino data. While we are
mostly interested in a LR symmetric setup with a gauge group
$SU(3)_C\times SU(2)_L \times SU(2)_R \times U(1)_{B-L}$, most of the
discussion in this section applies also to inverse seesaw within the
SM. We will formulate this section in the LR context and we will
comment on differences between inverse seesaw within the LR symmetric
and the SM scenarios at the end of the section.

\subsection{Setup} 
\label{subsect:Setup}


We work in the basis in which the mass matrix of the charged lepton is
diagonal, with the $e,\mu,\tau$ flavors identified by the mass
eigenvalues.  We write the inverse seesaw mass matrix in the
interaction basis for the neutral states
${\cal N} = \left( \nu_L, N^c_R, S^c_R\right)^T $ where
$\nu_L = \left(\nu_e,\nu_\mu,\nu_\tau\right)^T$ is the vector of the
$SU(2)$ partners of the LH charged leptons containing the (mainly
light eigenstate) LH neutrinos, $N_R=\left(N_e,N_\mu,N_\tau\right)^T$
is the vector of the neutral member of the $SU(2)_R$ doublets
$\ell_R=(N_R, e_R)$, and $S_R=\left(S_1,S_2,S_3\right)^T$ is a vector
of gauge singlet fermions for which a Majorana mass term
$\mu \overline{S^c} S$ is allowed.  In $3\times 3$ block notation the
mass matrix reads:
\beq
\label{eq:M}
{\cal M} = 
\begin{pmatrix}
0 & m_D^T & 0 \cr
m_D & 0  & M_R \cr
0 & M_R^T & \hat\mu
\end{pmatrix}\,, 
\eeq
where the Majorana sub-matrix $\hat\mu$ (as well as the full
$\mathcal{M}$) is complex symmetric.  Any complex symmetric matrix $m$
of any dimension can be factorized in a unique way as
$m=W^*\hat m W^\dagger$ where $\hat m$ is diagonal with real and
positive eigenvalues,  and $W$ is unitary.  Then, by redefining the
gauge singlets $S$ via a unitary rotation $W(\mu)$ we can always bring
$\mu$ into diagonal form $\hat \mu$ as is implicit in \eq{eq:M}.  As
regards $M_R$, if the fields $N_R$ were unrelated to the SM leptons
further field redefinitions would be possible.  However, in the LR
model the $N_R$'s  sit in the same $SU(2)_R$ multiplets with the RH
SM leptons, and once a redefinition of $\ell_R$ (together with a
redefinition of $\ell_L$) is used to bring into diagonal form the
charged lepton mass matrix, the only residual freedom is in three
vectorlike phase redefinitions of $\ell_{L,R}$ proportional to the
three diagonal $U(3)$ generators $I,\lambda_3,\lambda_8$ which commute
with the diagonal mass matrix. This can be used to remove three phases
from $M_R$ which remains otherwise generic with $9+6$ (real +
imaginary) parameters.  Finally, because of LR symmetry in exchanging
the L and R labels, the complex matrix $m_D$ is  symmetric.

Exact diagonalization of the mass matrix \eq{eq:M} can be 
performed via a transformation of the field basis with 
a  unitary matrix $\mathcal{V}$ such that 
\beq
\label{eq:Msymm}
\hat{\mathcal{M}}  =  \mathcal{V}^T \,  \mathcal{M} \, \mathcal{V} \,, 
\eeq
is diagonal.  Of course, in the general case this can only be done
numerically (our numerical study indeed relies on a precise numerical
diagonalization of the full $9\times 9$ matrix).  However, assuming
that the three sub-matrices in \eq{eq:M} have mass scales arranged
hierarchically $\mu,\, m_D \ll M_R$, an approximate diagonalization
can be performed in analytic form yielding:
\beq
\label{eq:Mapp}
\hat{\mathcal{M}}' = {\mathcal{V}'}^T \, \mathcal{M} \, \mathcal{V}' \approx \hat{\mathcal{M}}
\eeq
where $\mathcal{V}' \approx \mathcal{V}$ is non-unitary by terms of
$\mathcal{O}(m_D/M_R)$ (we denote with a prime non-unitary
transformation matrices, as well as mass matrices obtained via
non-unitary transformations).  Clearly $\hat{\mathcal{M}}'$ deviates
from exact diagonal form: terms of $\mathcal{O}(\mu\, m_D/M_R)$ will
appear in the non-diagonal entries coupling the light and heavy
sectors, and terms of $\mathcal{O}(\mu)$ will appear in the non
diagonal entries of the heavy sector.  Below we give a brief
description of this approximate diagonalization procedure, which will
also be useful to establish notations.

\subsection{Stepwise approximate diagonalization} 
\label{subsect:Appr}

Approximate diagonalization can be carried out in four steps.  The
first step is to bring $M_R$ into diagonal form.  Let us decompose
$M_R$ in terms of two unitary matrices $U_R,\,V_R$ and a diagonal
matrix of mass eigenvalues $\hat M_R$:
\begin{eqnarray}
\label{eq:bidiMR}
M_R = U_R{\hat M}_RV_R^{\dagger} \,.  
\end{eqnarray}
As we have remarked above, $M_R$ contains nine real and six imaginary
parameters. Then, by matching the number of parameters between the LH
and RH sides of \eq{eq:bidiMR} we see that $U_R$ and $V_R$ can be taken
as special unitary, with three real angles and three phases each.  The
matrix $U_R$ is an important quantity since, for example, it will appear
in the RH charged currents coupling $N_R$ to the charged leptons.  By
defining a block-diagonal matrix
$V_1={\rm diag}\left(\mathbb{I}_3,U_R^*,V_R\right)$, where
$\mathbb{I}_3$ is the $3\times3$ identity matrix, it is easy too see
that in the matrix $\mathcal{M}_1=V_1^T \mathcal{M} V_1$ an exact
diagonalization $M_R\to \hat M_R$ is obtained, while at the same time
$\hat \mu \to \mu^V\equiv V_R^T \hat \mu V_R$ and the entries $m_D$
($m_D^T$) get replaced by $D$ ($D^T$) defined as:
\beq
\label{eq:D}
D \equiv U^\dagger_R m_D\,. 
\eeq
The next step $\mathcal{M}_2=V_2^T \mathcal{M}_1 V_2$ with
$V_2 = \frac{1}{\sqrt{2}}{\rm diag}(\sqrt{2},\sigma_1-\sigma_3)\otimes
\mathbb{I}_3$
brings $\hat M_R$ to the block-diagonal (2,2) and (3,3) entries and
also adds to these entries small corrections of $\mathcal{O}(\mu^V)$.
The $D$ terms \eq{eq:D} remain in the first row
$\left(\mathcal{M}_2\right)_{1j}=
v_D^T=\frac{1}{\sqrt{2}}(0,-D^T,D^T)$
and first column $\left(\mathcal{M}_2\right)_{j1}= v_D$.  Let us note
that since $V_1$ and $V_2$ are both unitary, no approximation has been
made so far in $\mathcal{M}_2$. The next step requires suppressing the
off-diagonal entries of order $m_D$.  This is obtained with a matrix
$V'_3$ such that
$\left(V'_3\right)_{1j}=
w_D^\dagger=(\mathbb{I}_3,D^\dagger,D^\dagger)\frac{1}{\sqrt{2} \hat
  M_R}$,
$\left(V'_3\right)_{j1}=-w_D$ and
$\left(V'_3\right)_{jj}=\mathbb{I}_3$.  It can be easily checked that
$V'_3 {V'_3}^\dagger$ deviates from the identity by
$\mathcal{O}(m_D^2/M^2_R)$. With this rotation, the off-diagonal
light-heavy entries in $\mathcal{M}'_3={V'_3}^T \mathcal{M}_2 V'_3$ get
suppressed to $\mathcal{O}(\hat\mu D/M^2_R)$ which, in the seesaw
approximation, can be neglected.  We have thus singled out 
in the (1,1) block the light neutrino mass matrix $m_\nu$, which  
can now be expressed, as is customary, in terms of the initial matrices
in~\eq{eq:M} as:
\begin{equation}
\label{eq:mnueff}
  m_\nu \simeq m_D^T \frac{1}{M_R^T} \hat\mu\frac{1}{M_R}m_D \,. 
\end{equation}
We see from this equation that suppression of the light neutrino
masses can be obtained thanks to small values of $\hat\mu$, without
the need of exceedingly small values of $m_D/M_R$. This can allow for
$N_R$ to live at relatively low energy scales, possibly within
experimental reach. Being symmetric by construction, $m_\nu$ can be
diagonalized as
\begin{equation}
\label{eq:VL}
  \hat m_\nu = {V_L}^T\, m_\nu\, V_L\,, 
\end{equation}
with $V_L$ unitary.  Note that $V_{L}$ differs from the exact
(non-unitary) light neutrinos mixing matrix $V'_L$ by
$ {\cal O}(\frac{m_D}{M_R})$.  In our study we will neglect these
small terms and we will identify $V_L=V'_{L}$.
A last rotation, by means of the
unitary matrix
$V_4 = {\rm diag}\left( V_L,\, i\, \mathbb{I}_3,\,
  \mathbb{I}_3\right)$,
can now be performed on $\mathcal{M}'_3$ to bring $m_{\nu}$ into
diagonal form (this also renders positive the heavy mass entries in
the (2,2) block that have acquired a negative sign).  Neglecting the
small off-diagonal entries, the final matrix
$\mathcal{M}' = V_4^T \mathcal{M}'_3 V_4$ reads:

\begin{equation}\label{eq:Mat4}
\mathcal{M}'\simeq 
\begin{pmatrix}
 {\hat m_{\nu}}  & 0 & 0 \\
 0 &   {\hat M_R}^{-} & 0\\
 0  &  0 &  {\hat M_R}^{+}
\end{pmatrix} .
\end{equation}
The eigenvalues of the two $3\times 3$ heavy-heavy blocks
${\hat M_R}^{\pm}$ receive corrections of
$\mathcal{O}(m_D^2/\hat M_R)$ after the $V_4$ rotation. However, these
corrections are the same for both blocks, so that they can be
conventionally absorbed into a common term $\hat M_R$. Instead,
contributions of order $\hat\mu$ appear with opposite sign, and this
is important because it generates small splittings between pairs of
heavy states.  For our analysis it is then sufficient to define the
heavy mass eigenvalues in \eq{eq:Mat4} as
${\hat M_R}^{\pm} = {\hat M}_R \pm \frac{1}{2} \mu^V$, keeping in mind
that they represent three pairs of almost degenerate (quasi-Dirac)
neutrinos with large masses $(\hat M_R)_{ii}$, split by three small
quantities
$(\Delta M)_{ii}=(\hat M^+_R)_{ii} - (\hat M^-_R)_{ii} = (\mu^V)_{ii}$
where $\mu^V \equiv V_R^T \hat\mu V_R^T$ (this last definition is
given here for the sake of precision, but being $V_R$ and $\hat \mu$
in any case arbitrary, in the following we will simply denote the mass
splittings generically as $\Delta M=\mu$).

\subsection{Couplings to the gauge bosons and to the Higgs} 
\label{subsect:cpl}

The approximate mixing matrix $\mathcal{V}'=V_1\,V_2\,V'_3\,V_4$ 
derived in the previous section controls
the structure of the couplings between the
LR gauge bosons and the mass
eigenstates. Its explicit form is:
\begin{equation}
  \label{eq:Vp}
  \mathcal{V}' = 
\begin{pmatrix}
V_L & 
\frac{i}{\sqrt{2}} \xi^\dagger  & \frac{1}{\sqrt{2}} \xi^\dagger \cr
0 & -\frac{i}{\sqrt{2}} U_R^* & \frac{1}{\sqrt{2}} U_R^*  \cr 
- V_R \xi  V_L & \frac{i}{\sqrt{2}} V_R & \frac{1}{\sqrt{2}} V_R  
\end{pmatrix} 
\end{equation}
where for convenience we have introduced the $3\times 3$ matrix 
of   small mixings:
\begin{equation}
  \label{eq:Xi}
\xi = \frac{1}{\hat M_R}\,D =   \frac{1}{\hat M_R}\,U_R^\dagger\,
m_D\,. 
\end{equation}
The derivation of the charged current (CC) couplings to $W_{L,R}^\pm$
and of the neutral current (NC) couplings to $Z_{L,R}$ is outlined
below. It is left understood that the known SM couplings fix the
normalization modulo a factor of the ratio of the gauge couplings
$g_R/g_L$.  Let us introduce a vector $E=(e_L,e^c_R,0)^T$ for the
left-handed (mass eigenstate) charged fermions, and recall that the
neutral states are arranged in another vector
$ \mathcal{N}=(\nu_L,N^c_R,S^c_R )^T$. The LH and RH charged currents
can be written (in two component notations) as:
\begin{eqnarray}
  \label{eq:CCL0}
  J^{-\mu}_L &=& \frac{1}{\sqrt{2}}\, E^\dagger \bar \sigma^\mu \, p_L\, \mathcal{N}\,,  \\
  \label{eq:CCR0}
  J^{-\mu}_R &=& \frac{1}{\sqrt{2}}\, E^\dagger  \bar \sigma^\mu \, p_R\, \mathcal{N}\,,
\end{eqnarray}
where $\bar \sigma^\mu= (1,-\vec \sigma)$ are the spinor matrices,
and $p_{L,R}$ are 
the projectors onto the neutral members of the L and R
multiplets corresponding to $9\times 9$ matrices  
which, in $3\times 3$ block notation, are given by
$(p_L)_{11}=\mathbb{I}_3$, $(p_R)_{22}=\mathbb{I}_3$ with zero in all
other entries. In the seesaw approximation, the neutral mass
eigenstates are related to the interaction eigenstates as
$ \mathcal{N} = {\mathcal{V}'}\, N$ with $N = (\nu,N_-,N_+)^T$,  where
$\nu$ represents the three light neutrinos
and $N_{\pm}$ correspond to the heavy neutrinos 
respectively with mass eigenvalues $M_R^\pm$.
Projecting onto the mass eigenstates and converting to the usual 
four-component spinor  notation for gauge currents we have:
\begin{eqnarray}
  \label{eq:CCL}
  J^{-\mu}_L  &=& 
\frac{1}{\sqrt{2}}\, \bar e_L \gamma^\mu\,  V_L \nu  
+ \frac{1}{2}  \; \bar e_L  \gamma^\mu \, \xi^\dagger\, (N_+ + i N_-) \,, 
\\
  J^{-\mu}_R  &\sim& 
 \frac{1}{2} \;  \overline{e_R^c}\;  \gamma^\mu\, U_R^*\,
                     (N_+ - i N_-) \,. 
\end{eqnarray}
NC couplings are also important since they can give rise to 
$N_\pm \to Z\nu$ decays. In the interaction basis the  NC
for the neutral states are:
\begin{eqnarray}
  \label{eq:NCL0}
  J^{0\mu}_L &=&  \frac{1}{2}\,\mathcal{N}^\dagger i \bar \sigma^\mu \, p_L\, \mathcal{N}\,,  \\
  \label{eq:NCR0}
  J^{0\mu}_R &=&  \frac{1}{2}\,\mathcal{N}^\dagger i \bar \sigma^\mu \, p_R\, \mathcal{N}\,,
\end{eqnarray}
which in the mass eigenstate basis yields:
\begin{eqnarray}
  \label{eq:NCL}
  J^{0\mu}_L &=& 
\frac{1}{2} \bar\nu \gamma^\mu  \, \nu +
\frac{1}{2\sqrt{2}} \left[\bar\nu \gamma^\mu V^\dagger_L \xi^\dagger (N_++i
                 N_-) + H.c. \right] \,, \\
  \label{eq:NCR}
  J^{0\mu}_R 
&=& 
\frac{1}{4}\, 
(\bar N_+ +i \bar N_-) \gamma^\mu  (N_+-i N_-)\,.  
\end{eqnarray}
In the first equation  we have neglected additional terms involving
$N$-$N$ couplings which are suppressed as $\xi\xi^\dagger$. 
As can be seen from the second equation, in the approximation in which
terms of order 
$ \mu/{\hat M_R} \xi$ 
are neglected there are no R-handed
neutral currents between heavy and light  neutrinos.  Finally, the
fermion-scalar coupling $\frac{1}{v} N_R^\dagger m_D \nu_L\, H$ gives
the following interactions between the heavy $N_\pm$'s, the Higgs and
the light neutrinos:
\begin{eqnarray}
\label{eq:Hcouplings}
{\mathcal L}_H 
&=& \frac{1}{\sqrt{2}}  (\bar N_++i \bar N_-)\left[ U_R^T
\frac{m_D}{v}\, V_L \right] \nu\, H \\ 
\nonumber 
 &+&  \frac{1}{2}
 (\bar N_++i \bar N_-)\left[U_R^T \frac{m_D }{v} \xi^\dagger \right]  (N_++i
 N_-) H  + H. c. \,.
\end{eqnarray}

\subsection{A useful parametrization of the inverse seesaw in LR models}
\label{subsect:CI}

In~\cite{Casas:2001sr} a clever parametrization of the Dirac mass
matrix of the type I seesaw was put forth, and it is referred to as
the Casas-Ibarra (CI) parametrization.  In this parametrization $m_D$
is expressed in terms of low energy observables (light neutrino mass
eigenvalues and mixing angles), of the seesaw heavy mass eigenvalues,
and of an arbitrary complex orthogonal matrix $\mathcal{R}$.  One of
the most useful features of the CI parametrization is that it allows
to generate random samples of $m_D$ which by construction reproduce
all the low energy data, which is a quite valuable property when one
wants to scan over the model parameter space.  As we detail below, also
for the inverse seesaw in LR models it is possible to introduce a
parametrization that has analogous properties, namely that allows 
to scan over the unknown physical masses and couplings
($U_R,\,V_R,\, m_D,\, M_R,\,\hat\mu$) while automatically reproducing
all the low energy data.

Let us start by writing the light neutrino mass matrix in diagonal
form (see \eq{eq:mnueff} and \eq{eq:VL}):
\begin{eqnarray}
\label{eq:hatmnueff}
  \hat m_\nu &=&  
V_L^T m_D^T \frac{1}{M_R^T} \hat\mu\frac{1}{M_R}m_D V_L \,.
\end{eqnarray}
Let us now write $m_D$ as: 
\begin{eqnarray}
  \label{eq:mD}
m_D &=&  M_R \,
\frac{1}{\sqrt{\hat\mu}}\, {\cal R} \, 
\sqrt{\hat m} \, V_L^\dagger\,. 
\end{eqnarray}
By inserting \eq{eq:mD} into the RH side of \eq{eq:hatmnueff} (or by
extracting directly $\mathcal{R}$ from \eq{eq:mD}) it can be verified
that $\mathcal{R}$ must satisfy the condition
$\mathcal{R}\mathcal{R}^T=\mathcal{R}^T\mathcal{R}=I$, but is otherwise
arbitrary, and thus it can be written  as a generic $3\times 3$
orthogonal matrix in terms of three complex angles.
Rewriting $M_R$ in the previous equation 
according to \eq{eq:bidiMR} we obtain  
\begin{equation}
  \label{eq:Ddef}
D=  U_R^\dagger\,m_D =  \hat M_R \, V_R^\dagger \,
\frac{1}{\sqrt{\hat\mu}}\, {\cal R}\, 
\sqrt{\hat m} \, V_L^\dagger\,.
\end{equation}
The RH side of this equation is written in terms of the low energy
observables ($\sqrt{\hat m} \, V_L^\dagger$) while the other quantities
are arbitrary.  The crucial point now is to factor the generic
$3\times 3$ complex matrix $D$ as defined in \eq{eq:Ddef} into a
unitary matrix ($U_R^\dagger$) and a symmetric matrix ($m_D$).
This can be achieved by factorizing $D$ in its singular value
decomposition (SVD) in terms of two unitary matrices $W$ and $Q$ and a real
diagonal matrix with non-negative entries $\hat D$:
\begin{equation}
  \label{eq:Dsing}
  D = W \cdot\hat D \cdot Q^\dagger =  (WQ^T)\cdot (Q^*\hat D
  Q^\dagger) \equiv    \tilde U_R^\dagger    \tilde m_D
\,, 
\end{equation}
where, in the second step, we have inserted $Q^TQ^*=\mathbb{I}_3$ in
order to build up a unitary matrix $\tilde U_R$ and the symmetric
matrix $\tilde m_D$.  However, $\tilde U_R$ and $\tilde m_D$ found in
this way are just one among a threefold infinite class of
possibilities, spanned by the freedom in switching phases between
$\tilde U_R$ and $\tilde m_D$ (all the moduli are instead uniquely
fixed).  This is due to the fact that the SVD decomposition is not
unique, since there are 9 phases in $D$ and $12$ in its decomposition
in terms of $W,\hat D$ and $Q$.  However, as discussed
below \eq{eq:bidiMR}, without loss of generality $U_R$ can be taken
special unitary with just $3$ phases, and doing so the counting of
parameters between the LH and RH sides of \eq{eq:Dsing} matches.  Let
us then introduce a diagonal matrix of phases
$\Phi = \mathrm{diag}(e^{i \varphi_1}, e^{i \varphi_2},e^{i  \varphi_3})$
and make the identification    
\begin{equation}
\label{eq:URpar}
   U_R = Q^* \Phi^* W^\dagger\,,  \qquad 
m_D = Q^* \Phi^*\hat D Q^\dagger \,,
\end{equation}
which clearly preserves $U_R^\dagger m_D=D$ and the symmetric nature
of $m_D$.  The values of $\varphi_i$ can then be fixed to achieve the
desired form for $U_R$.  Therefore, in the LR inverse-seesaw, given
for example a set of RH neutrino masses $\hat M_R$ and of LNV
parameters $\hat \mu$ of specific interest, the parametrization
\eq{eq:Dsing} together with \eq{eq:URpar} yields both $m_D$ and $U_R$
in terms of two arbitrary matrices: a complex orthogonal matrix
$\cal R$ and a special unitary matrix $V_R$ with just three phases,
while, by construction, all the low energy neutrino data are
automatically reproduced.


The discussion in this section assumed an inverse seesaw within the
left-right symmetric group. However, it is straightforward to adapt most
our discussion to inverse seesaw models with the same block structure
of $\mathcal{M}$ as in \eq{eq:M}, but for which $N_R$ is not related
to $\ell_R$, i.e. the standard model gauge group.  In this case $m_D$
is not constrained to be symmetric and we gain the freedom of
redefining $N_R$ via a $U(3)$ transformation. This allows to reabsorb
$U_R$ defined in~\eq{eq:bidiMR} via a field rotation, while $V_R$
remains defined in terms of three real and three imaginary parameters.
Then $U^\dagger_R$ can be simply dropped from~\eq{eq:Ddef} whereas
$D=m_D$ remains generic. \footnote{Of course, within the 
SM group there are no right-handed gauge interactions, 
see section \ref{subsect:cpl}.}


\section{Opposite sign to same sign dilepton ratio}
\label{sect:Rratio}

In this section we estimate the ratio of production of pairs of leptons
with the same sign and we compare it with the rate of production of
pairs of leptons of opposite sign.  The ratio between these two
observables is denoted as $R_{ll}$. In both cases the production rates
are dominated by processes with on-shell (or nearly on-shell) $N_R$'s
and therefore, under the natural assumption that the mass splitting
between the different pairs is large (we typically expect
$M_{Rj}^\pm - M_{Rk}^\pm \sim \mathcal{O}(M_R)$), it is sufficient to
study just a single pair of quasi-Dirac $N_\pm$.  SS dilepton
production occurs for example through the LNV process
$\bar q q \to W_R^+ \to \ell^+_\alpha N_\pm \to \ell^+_\alpha
\ell^+_\beta W_R^*$,
where $(\bar q) q$ denote (anti-)quark partons inside the colliding
protons, $N_+$ and $N_-$ are the two heavy neutrinos mass eigenstates,
$W_R^*$ is an off-shell RH gauge boson that will eventually decay
dominantly in two jets, and $\ell_\alpha,\, \ell_\beta$ are two
leptons not necessarily of the same flavor. Opposite sign pairs of
leptons can be produced via the LN conserving process
$\bar q q\to W_R^+ \to \ell^+_\alpha N_\pm \to \ell^+_\alpha
\ell^-_\beta W_R^*$.
Clearly, in order to produce the $N_\pm$ intermediate states on-shell
via the decay of an on-shell $W_R$, $M_{W_R} > M_R^\pm$ is
required. We further assume $M_{W_R} \not \gg M_R^\pm$ so that the
$N_\pm$ mass eigenstates can be treated in the non-relativistic
approximation.

Before entering into details let us try to figure out qualitatively
what type of result we can expect.  When the on-shell $W^+_R$ decays,
an $\ell^+$ anti-lepton is produced together with a heavy neutrino of
$\ell$-flavor $N_\ell$, which corresponds to a coherent superposition
of the two mass eigenstates $N^\pm$. Given that the same decay
channels are open for both $N^\pm$, the time-evolution of the initial
$N_\ell$ will be characterized by a typical oscillating behavior with
frequency $\Delta M = M^+-M^- = \mu$. There is another important
scale in the problem, that is the $N^\pm$ lifetime
$\tau = 1/\Gamma$.\footnote{Since $N^\pm$ have the same decay
  channels, and only a tiny mass difference, we expect for the width
  difference $\Delta \Gamma = \Gamma^+ - \Gamma^- \ll \Delta M$ so
  that $\Delta \Gamma$ is always negligible. This is analogous to what
  happens in the $B^0-\bar B^0$ meson system (see
  e.g. ref.~\cite{Nir:1992uv}).}  If $\Delta M \gg \Gamma$ the
lifetime is long enough that complete separation of the $N^\pm$ wave
packets can occur.  Coherence between the two mass eigenstates is
completely lost before the decays, and decays will then proceed as in
the usual Majorana case, yielding equal probabilities for SS and OS
dileptons events, i.e. $R_{ll}=1$.  (Ideally, in this situation we can
imagine that the mass of the intermediate state can be reconstructed
from the invariant mass of the $N$ decay products $m_{\ell_2 jj}$ to
be $M^+$ or $M^-$, in which case the above result is obvious.)  In the
opposite limit $\Delta M \ll \Gamma$ decays occur at a time
$t_D \sim \tau \ll 1/\Delta M$, that is before the onset of
oscillation effects, so that $N_\ell(t_D) \approx N_\ell(0)$.  In this
case only the LN conserving transition $N_\ell(t_D) \to \ell^-$ can
occur and $R_{ll}=0$.  Namely, when the $N^\pm$ mass degeneracy (in
units of $\Gamma$) is sufficiently strong, the pure Dirac case is
approached.  It is then clear that the interesting regime occurs when
the oscillation frequency is of the order of the lifetime, viz when
$\mu =\Delta M \approx \Gamma$. Only in this case we can expect
$R_{ll} \neq 0, 1$.

From \eq{eq:Vp} we can write the $N_\ell$ heavy state produced in the
decay $W^+_R \to \bar\ell N_\ell$ and its conjugate state
$N_{\bar\ell}$ produced in the decay $W^-_R \to \ell N_{\bar\ell}$ in terms of
the mass eigenstates as:\footnote{One remark is in order: in the
  presence of CP violating effects, the modulus of the ratio of the
  two coefficients in the linear combinations
  \eqs{eq:NbarN1}{eq:NbarN2} can deviate from unity (CP violation in
  mixing \cite{Nir:1992uv}).  In the regime $\mu \sim \Gamma$ this
  type of CP violation can get resonantly enhanced, and in principle
  observable effects on the ratio $R_{ll}$ could be possible.  We neglect
  this possibility in our treatment.}
\begin{eqnarray}
  \label{eq:NbarN1}
  N_\ell &=& \frac{1}{\sqrt{2}} (N_+ - i N_-)\,, \\  
  \label{eq:NbarN2}
  N_{\bar\ell} &=& \frac{1}{\sqrt{2}} (N_+ + i N_-) \,. 
\end{eqnarray}
In writing these linear combinations we have neglected for convenience
the flavor mixing matrices $U_R$ (see \eq{eq:Vp}) since the products of
their matrix elements appearing in the LN conserving and
LNV amplitudes cancels in the ratio $R_{ll}$. However,  it should be
kept in mind that 
these matrix elements control the flavor
composition of both the SS and OS dilepton final states $\ell_i\ell_j$, 
and we reiterate that for generic mixing structures, $i\neq j$ events
have no reason to be suppressed with respect to $i=j$ events.

After a time $t$, the states in \eq{eq:NbarN} have evolved into 
\cite{Nir:1992uv}
\begin{eqnarray}
  \label{eq:gpm}
  N_\ell(t) &=& g_+(t) N_\ell + g_-(t)  N_{\bar\ell}\,, \\  
  \label{eq:gmp}
  N_{\bar\ell} (t)&=& g_-(t) N_\ell + g_+(t) N_{\bar\ell} \,, 
\end{eqnarray}
where the oscillating amplitudes read 
\begin{eqnarray}
  \label{eq:NbarN}
  g_+(t) &=& e^{-i M t}e^{-\frac{\Gamma}{2} t} \cos \left(\frac{\Delta M}{2}t \right)\,, \\  
  g_-(t) &=& i \,e^{-i M t}e^{-\frac{\Gamma}{2} t} \sin \left(\frac{\Delta M}{2} t\right)\,,   
\end{eqnarray}
with $M=\frac{1}{2}(M_++M_-)$ and, according to the discussion above,
we have neglected the effects of $\Delta\Gamma$.  Since the typical
heavy neutrino widths are too large to allow observing displaced
vertices (see next section), individual oscillation patterns cannot be
resolved. The SS to OS ratio $R_{ll}$ is then given by the ratio of
the time-integrated amplitudes squared (note that they include the time 
dependent weight factor of the heavy neutrinos lifetime):
\begin{equation}
  \label{eq:RatioR}
R_{ll} = \frac
{\int_0^\infty \left|g_-\right|^2 dt}  
{\int_0^\infty\left|g_+\right|^2 dt}  = \frac{ \Delta M^2}{2 \Gamma^2 +
  \Delta M^2} \,. 
\end{equation}
%
This result correctly reproduces the limiting cases discussed at the
beginning of this section, that is $R_{ll}\to 1$ as $\Gamma /\Delta
M\to 0$ (limiting Majorana case) and $R_{ll}\to 0$ as $(\Gamma/\Delta
M)^{-1} \to 0$ (limiting Dirac case).\footnote{This result disagrees
  with eq.(7) of ref.~\cite{Dev:2015pga} which displays an explicit
  dependence of $R_{ll}$ on the heavy neutrino mass $M$.}

\section{LHC Phenomenology}
\label{sect:Pheno}

In searching for heavy RH neutrinos within the framework of LR
symmetric models, both the ATLAS \cite{ATLAS:2012ak,Aad:2015xaa} and
the CMS collaboration \cite{CMS:2012uwa,Khachatryan:2014dka} assume
that the heavy neutrino decays proceed via an off-shell $W_R$ bosons,
with a branching ratio of 100\% for the decay mode $N \to l^{\pm}jj$
where $l$ represents a charged lepton of any flavor and $N$ represents
a generic heavy neutrino.  While this is a reasonable expectation for
LR models with an ordinary seesaw mechanism, the situation is very
different in models based on the inverse seesaw.  In our
framework in fact all the following decay modes can occur, and all with
sizeable branching ratios:
\begin{eqnarray}\label{eq:ndec}
N \to W_L^{\pm} + l^{\pm}\,, &&  N \to Z_L + \nu\,, 
\hskip5mm N \to h + \nu\,,  \\  
\nonumber
N \to (W_R)^{*} + l^{\pm} \to jj l^{\pm}
& , & 
N \to (Z_R)^{*} + \nu 
\to  (jj\; {\rm or}\; l^{+} l^{-})\nu \,, 
\end {eqnarray}
where $W_L$ and $Z_L$ are the (mostly) SM gauge bosons, $h$ is the SM
Higgs with mass $m_{h} \simeq 125$ GeV, and $\nu$ represents a light
neutrino of any flavor. In our analysis we also assume
$m_{N} < m_{W_R}$, where $m_N$ denotes collectively the  pair of mass
eigenvalues $(M_R^\pm)_{11}$ for the lightest heavy neutrinos, so that
the RH gauge bosons $(W_R)^{*}$ and $(Z_R)^*$ from $N=N_{1\pm}$ decays
are off-shell.  We also assume for simplicity
$(M_R^\pm)_{ii} > m_{W_R}$ for $i>1$ so that a single pair of RH
neutrinos contributes to the signal (this second assumption is not
necessary whenever the different pairs of heavy neutrinos are
sufficiently separated in mass so that the different invariant masses
of the decay products can be reconstructed with good confidence).  In the
numerical analysis we have also included the decay mode
$N \to (Z_R)^{*} + \nu$ although its branching is seesaw suppressed,
and therefore largely irrelevant with respect to the other decays (see
the comment below \eq{eq:NCR}).  In addition to the decay modes shown
in eq. (\ref{eq:ndec}), decays into additional scalars besides the
Higgs could also be possible, if they are lighter than $N$.  This
however, depends on unknown details of the scalar sector. Therefore,
for definiteness we will assume that any new scalar is heavier than
$N$ so that the dominant decay modes are all listed in \eq{eq:ndec}.

We first present some examples of numerical results 
corresponding to some  fixed value of $m_{W_R}$ and of $m_N$. 
This is justified by the fact that detection of $lljj$ signals at the
LHC would imply that $m_{W_R}$ and at least one $m_{N_i}$ will be
measured.  In all the plots low energy neutrino data are kept fixed at
their best fit point values for a normally ordered hierarchical
spectrum (no qualitative differences arise for inverted hierarchies).
We start by showing results for some fixed arbitrary choice of the
matrices $V_R$ and ${\cal R}$ (see section \ref{subsect:CI}).

\begin{figure}
\centering
\includegraphics[scale=0.6]{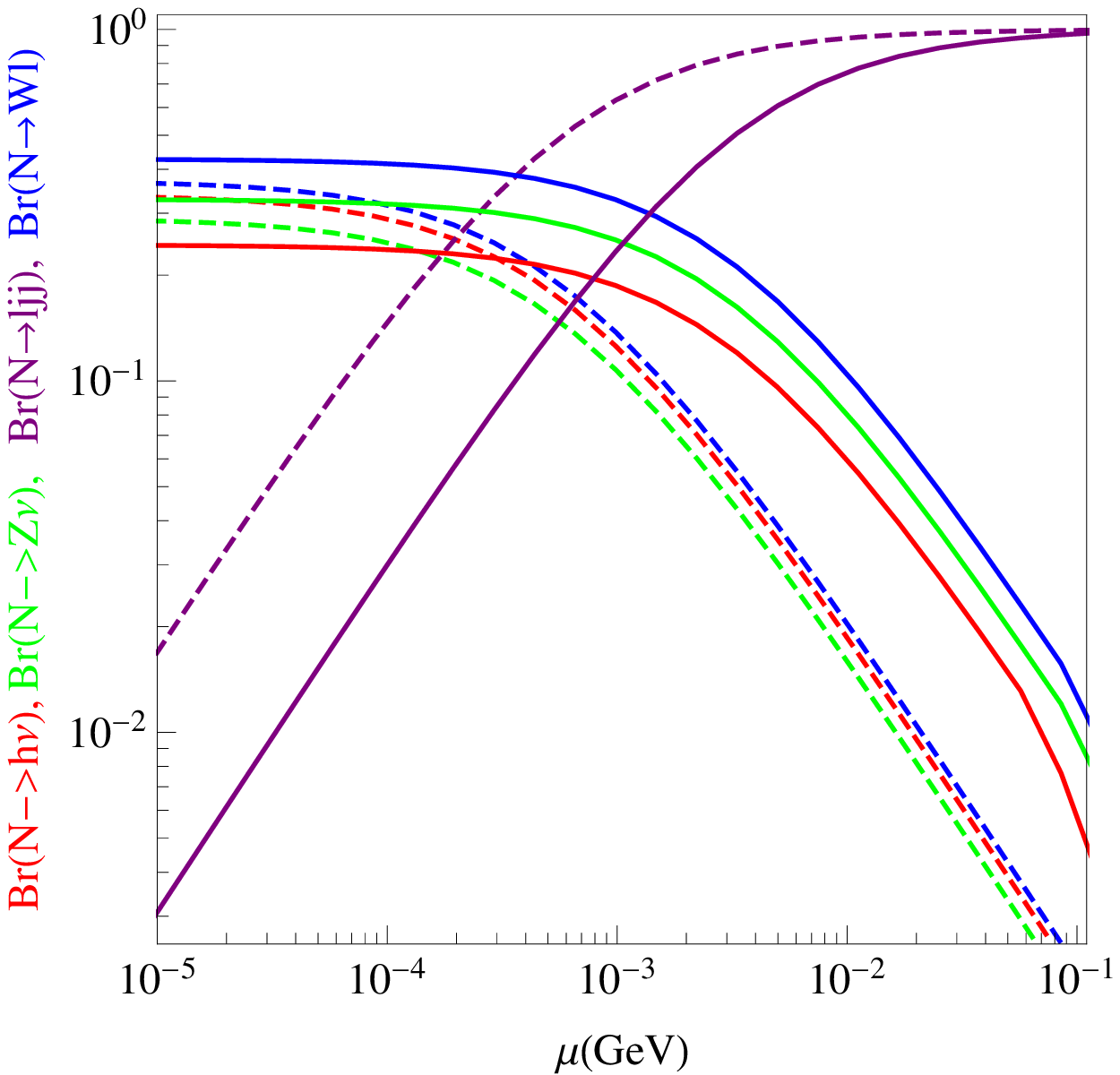}
\includegraphics[scale=0.6]{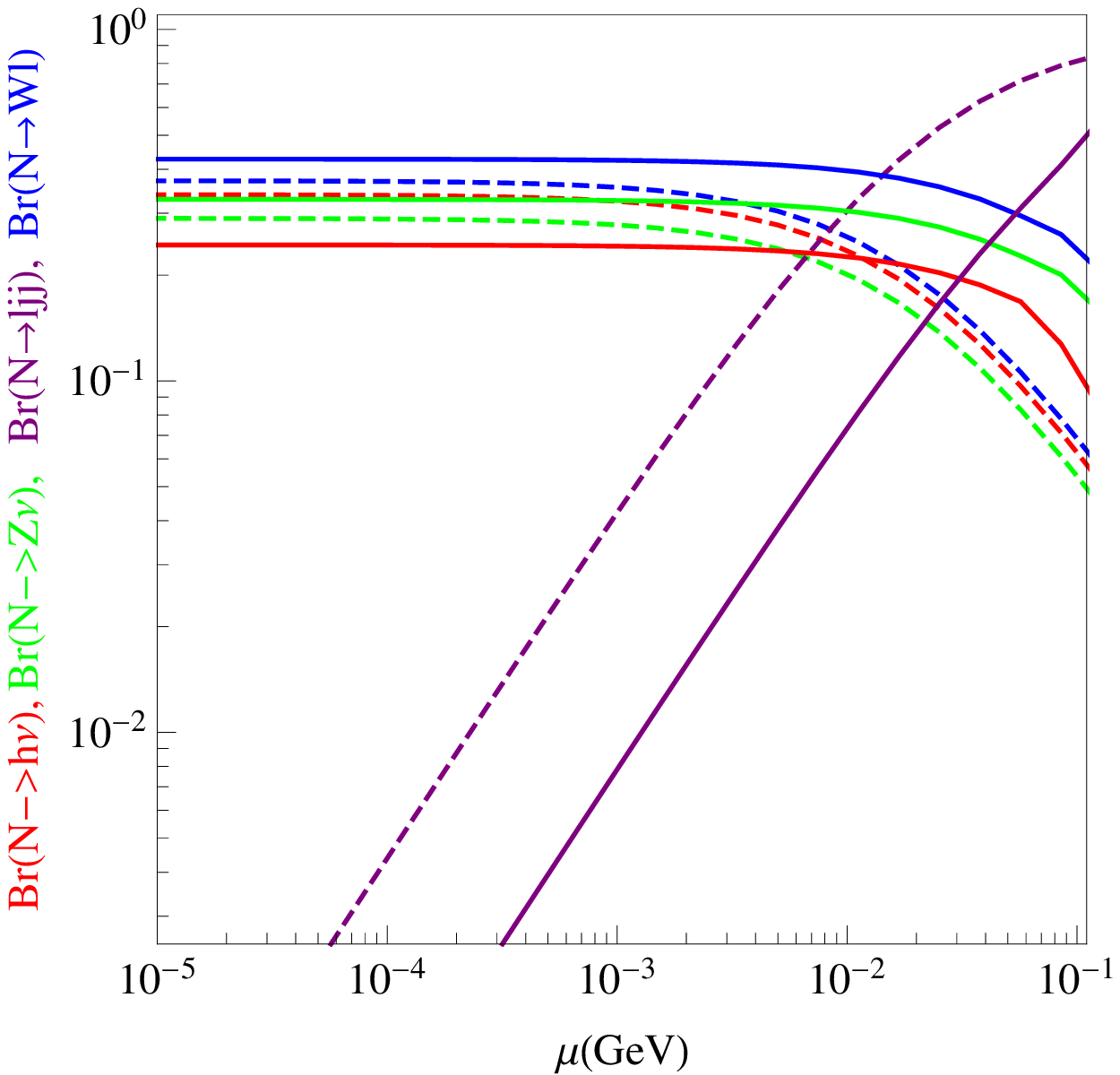}
\caption{\label{fig:BrVmu} Branching ratios for heavy neutrino decays
  as a function of $\mu$.  The blue lines are for $\Gamma(N\to W+l)$,
  green  for $\Gamma(N\to Z+\nu)$, red for $\Gamma(N\to h^0+\nu)$
  and purple for the three-body decay $\Gamma(N\to ljj)$. Solid lines
  correspond to $m_{N}=0.2$ TeV and dashed lines to $m_{N}=0.5$ TeV.
  The left panel is for $m_{W_R}=2$ TeV and the right panel for $m_{W_R}=5$
  TeV.  Lepton (and quark) final states are summed over flavor
  indices so that there is no dependence on fermion mixings.  }

\end{figure}

Fig.~\ref{fig:BrVmu} shows some typical values of the branching ratios
for different final states as a function of the LNV parameter $\mu$
ranging within the interval $[10^{-5},10^{-1}]$ GeV.  We have chosen
the two representative values $m_N=0.2\,$TeV (solid lines) and
$m_N=0.5\,$TeV (dashed lines), and two different values for the $W_R$
mass $m_{W_R}=2\,$TeV (left panel) and $m_{W_R}=5\,$TeV (right panel).
The lowest $m_{W_R}$ value corresponds roughly to the mass of the CMS
excess, while the largest one corresponds roughly to the maximum
$m_{W_R}$ that the LHC can probe in the next few years of running.  In
the final states we sum over the different quark and lepton
generations, so that the results are independent of neutrino
mixing. For small values of $\mu$, decays to SM gauge bosons dominate
the decay rates.  The branching ratios for
$N \to W_R^{*} + l^{\mp} \to l^{\mp}jj$ and for decays to SM gauge
bosons become similar for intermediate values of $\mu$, the detailed
ranges in which this occurs depend, however, rather strongly on the
values of $m_{N}$ and of $m_{W_R}$. For large values of $\mu$ three
body decays become dominant.  The qualitative behavior shown in
fig.~\ref{fig:BrVmu} can be understood from the equations presented in
the previous section.  In the inverse seesaw, the light neutrino
masses are given by eq. (\ref{eq:mnueff}). The equation contains the
three matrices $m_D$, $M_R$ and $\mu$ as free parameters. Keeping
fixed the light neutrino masses at values in agreement with the
experimental data and for fixed values of $M_R$, a scaling
$m_D \propto 1/\sqrt{\mu}$ is obtained.
Since all the couplings of the heavy neutrinos to SM gauge bosons are
proportional to $m_D$ (see the equations in section \ref{subsect:cpl}) 
decays to SM gauge bosons  dominate when $\mu$ is small.

\begin{figure}
\centering
\includegraphics[scale=0.6]{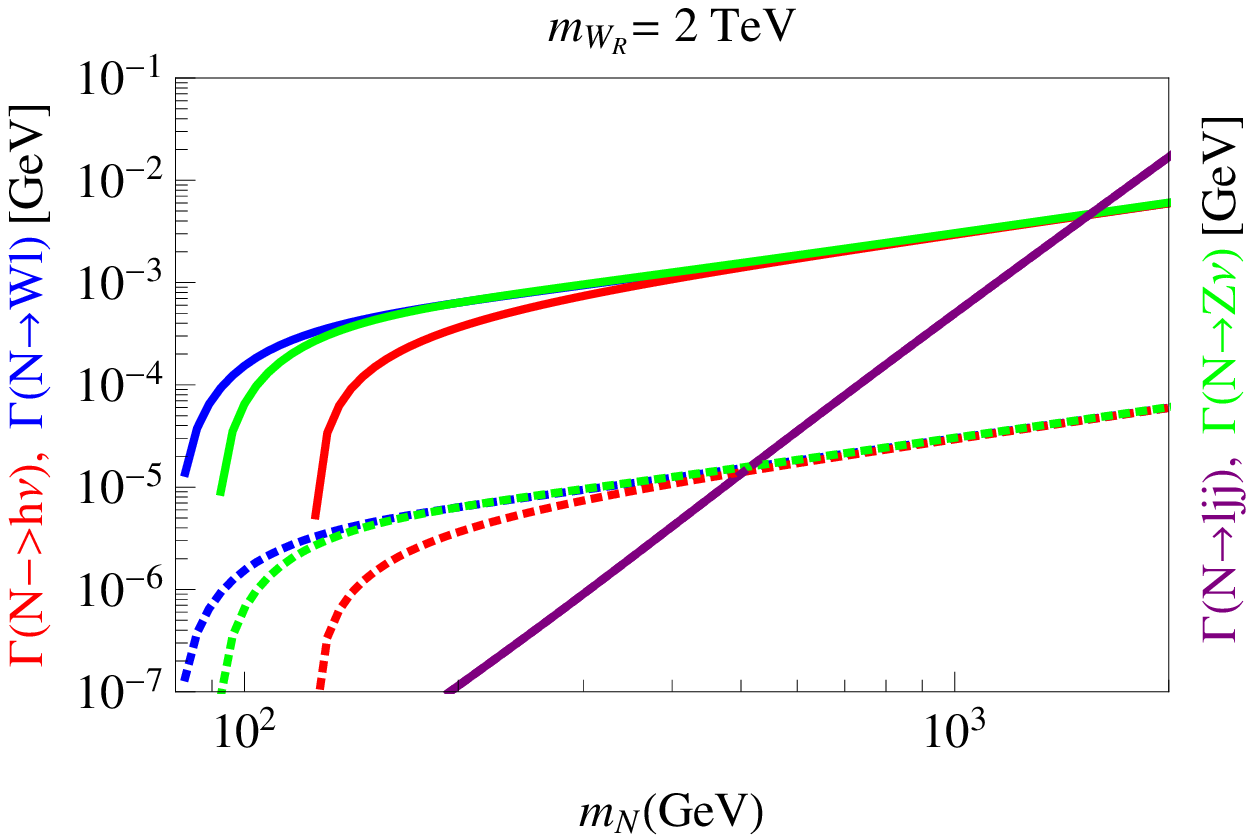}\hskip5mm
\includegraphics[scale=0.6]{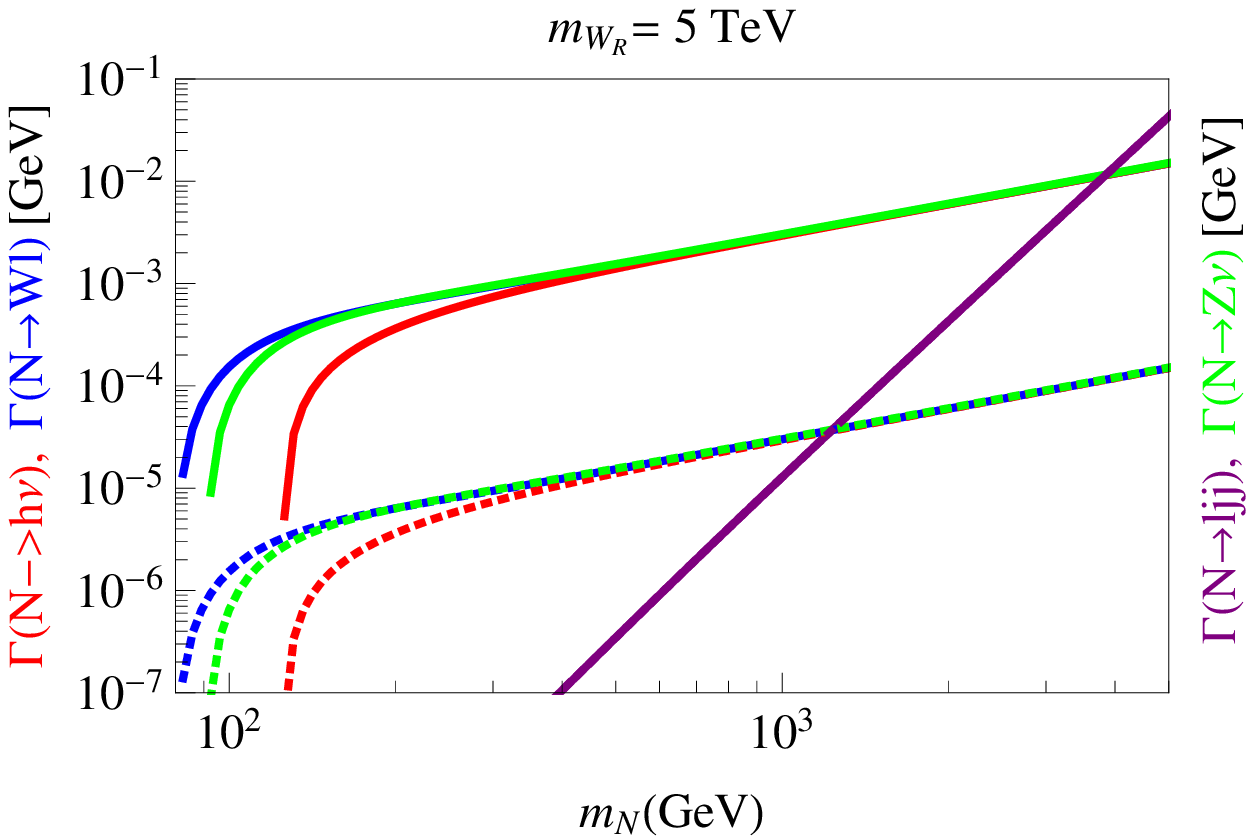}
\includegraphics[scale=0.6]{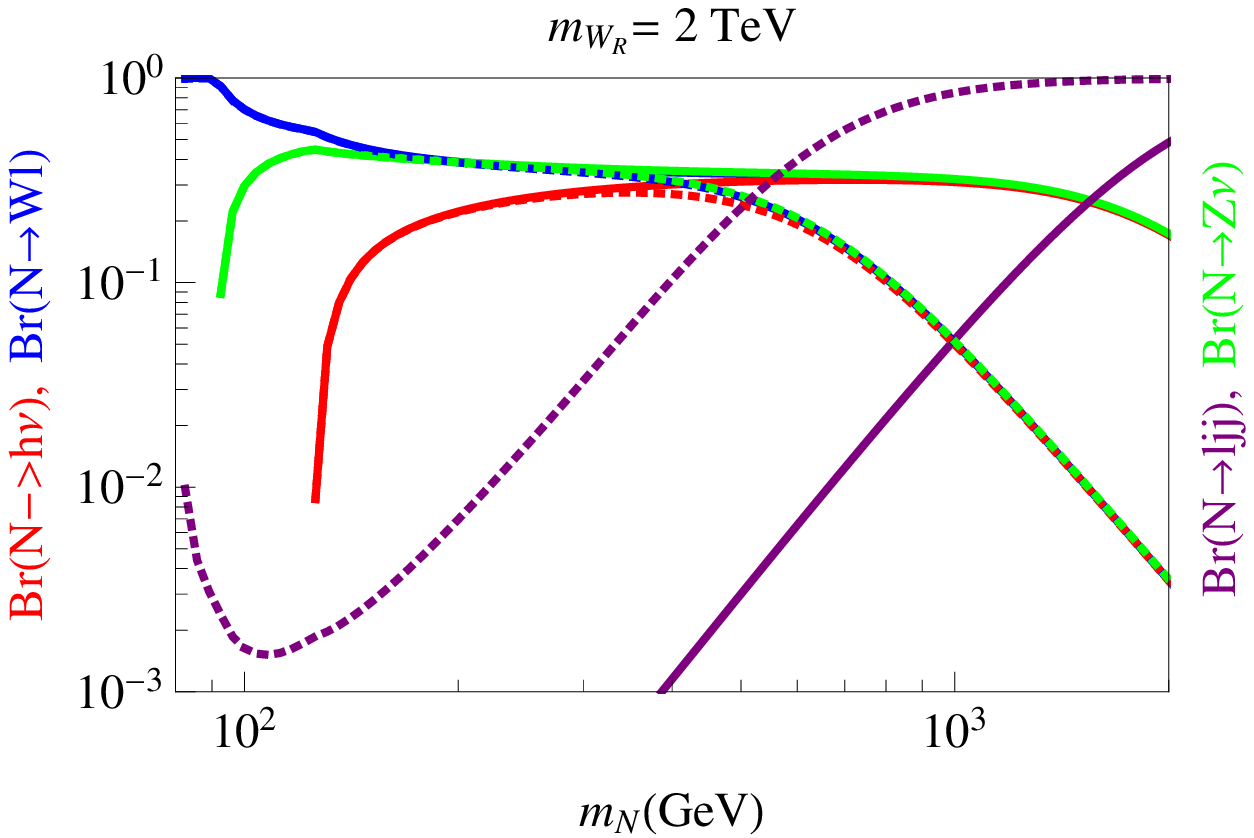}\hskip5mm
\includegraphics[scale=0.6]{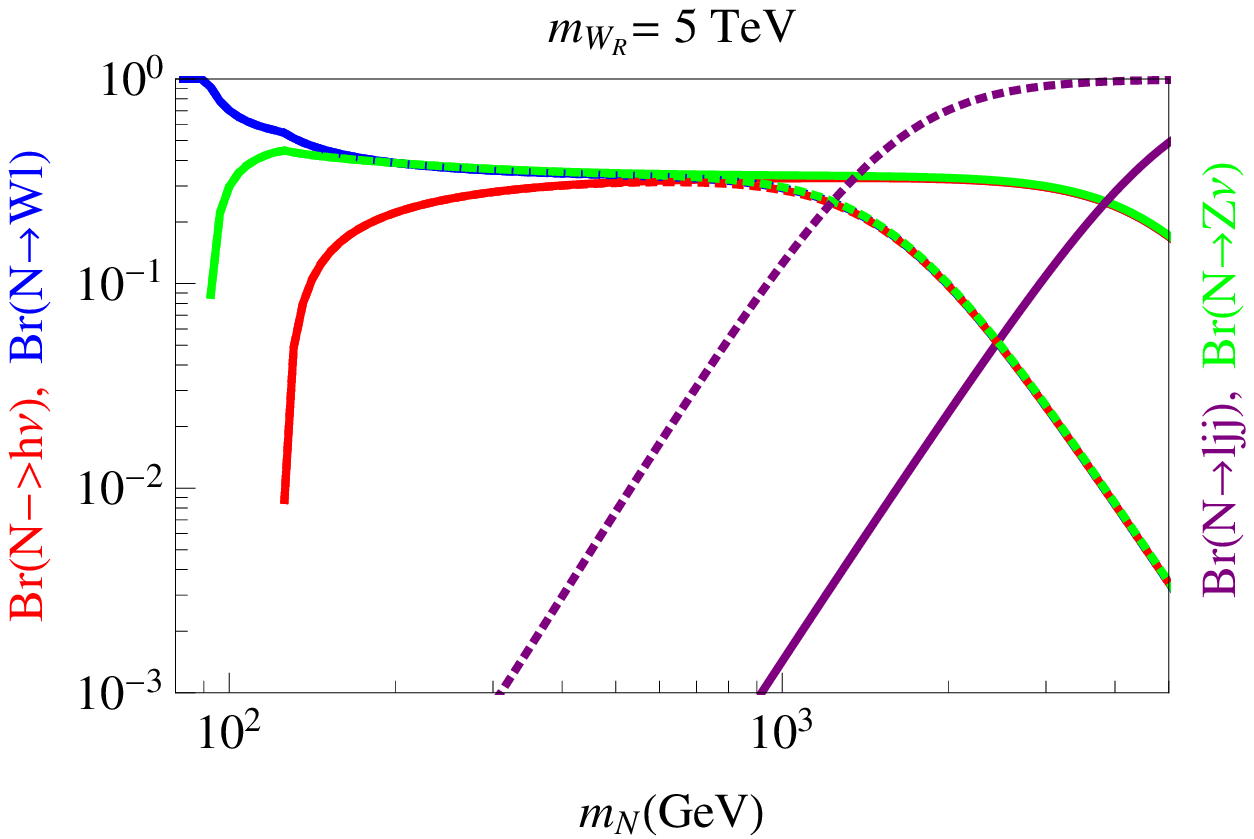}
\caption{\label{fig:Gami}Partial decay widths in GeV (top panel) and
  branching ratios (bottom panel) for $N$ decays. The blue, green and
  red lines are respectively for $\Gamma(N\to W+l)$,
  $\Gamma(N\to Z+\nu)$ and $\Gamma(N\to h^0+\nu)$ while the purple
  lines are for the three-body decay $\Gamma(N\to ljj)$. Solid lines
  correspond to $\mu=10^{-5}\,$GeV and dashed lines to
  $\mu=10^{-4}\,$GeV. Left panels correspond to  $m_{W_R}=2\,$TeV 
  and right panels to $m_{W_R}=5\,$TeV.  }
\end{figure}

Fig.~\ref{fig:Gami} shows the partial widths and branching ratios for
$N$ decays as a function of $m_{N}$ for the two values
$\mu=10^{-5}\,$GeV (solid lines) and $\mu=10^{-4}\,$GeV (dashed
lines).  Typical widths are in the range of
$\Gamma \simeq [10^{-7},10^{-2}]$ GeV, much too small to be directly
measured at the LHC, and too large to produce a displaced vertex.  For
small values of $m_N$, $N\to W_L l^\pm$ decays dominate the other
two-body decays. However, it is important to notice that for
$m_{N} \gg m_{h}$ the branching ratios of $N$ decays to $W_L$, $Z_L$
and $h$ summed over light flavors become all equal.  This can allow
to infer the branching ratio for $N$ decays to $W_L+Z_L+h$ from
the measurement of Br($N \to W^{\pm} + \sum_\alpha l^{\mp}_\alpha$)
alone.

Note also that the $W_L$ gauge bosons decay to jets with a branching
ratio of about $2/3 < 1$, and that $Z_L$ and $h$ do not lead to $ljj$
final states. This implies a reduction in the number of expected
$lljj$ events.  In the extreme case of very small $\mu$ and for
$m_{N} \gg m_{h}$, when the decays into SM bosons dominates, only
$1/9$ of the total number of decays are into $lljj$ final states
occurring mainly via the $N \to W_L + l \to ljj$ decay chain.  Let us
recall that experimental estimates are instead based on the assumption
that the only decay channel is $N\to W_R^* l^\pm$, implying that 100\%
of the decays correspond to $ljj$ final states.  Therefore, we can
expect that, within the present framework, the lower limit on
$m_{W_R}$ should be somewhat looser than the one quoted by the LHC
collaborations.  Let us also note that since  $W_L$'s are produced
on-shell, for $N \to W_L + l \to ljj$ decays, the invariant mass of the
jets should be peaked in correspondence to  $m_{W_L}$. Thus it
should be possible to separate kinematically these events from the
off-shell $W_R$ events. Such a measurement could be important to
establish large ``heavy-light'' mixing in the neutrino sector, that is
a general prediction of the inverse seesaw model.
Finally,  the fact that in the inverse seesaw models decays to SM
bosons can dominate in a wide region of parameter space is
again apparent also from  fig.~\ref{fig:Gami}.

\begin{figure}
\centering
\includegraphics[scale=0.6]{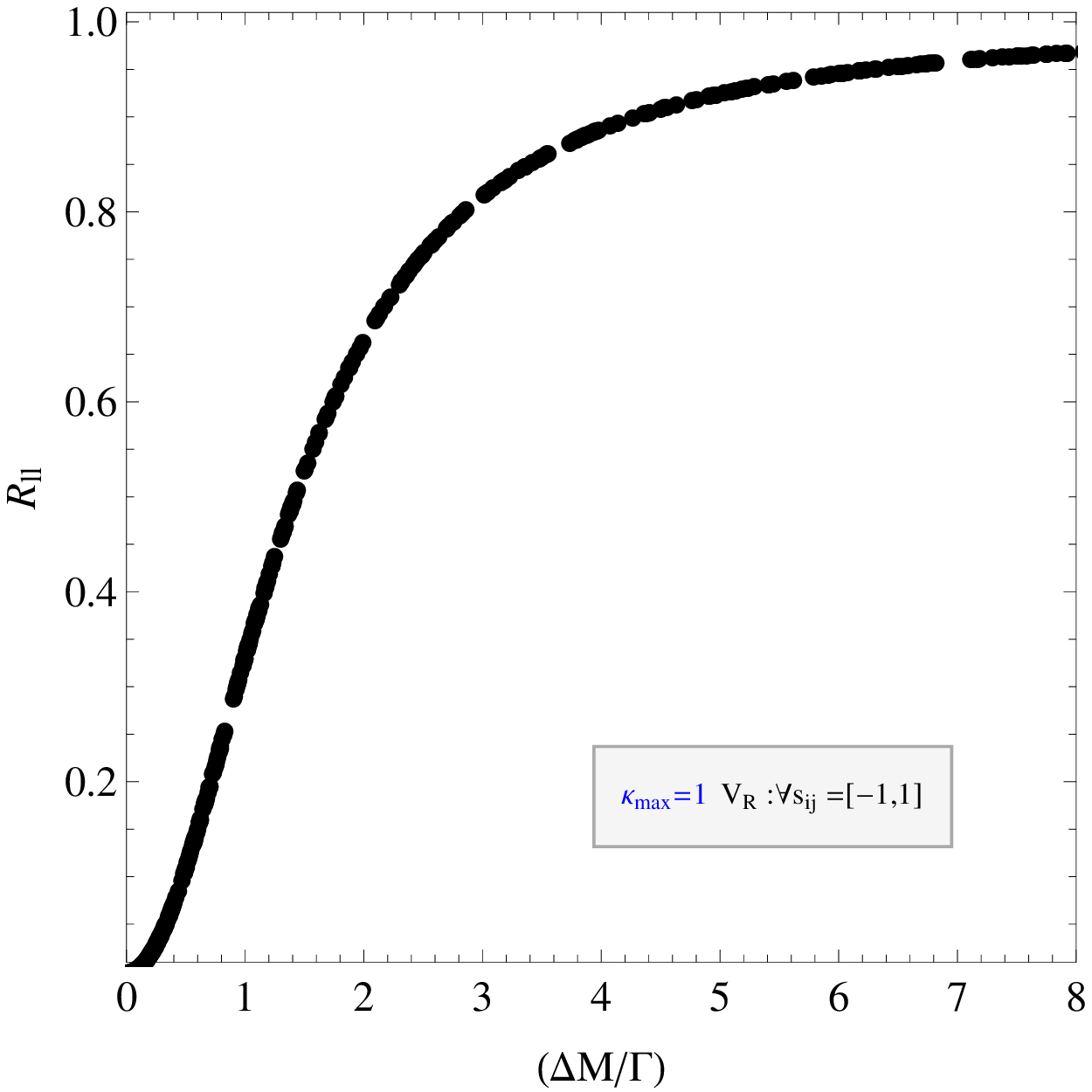}
\includegraphics[scale=0.6]{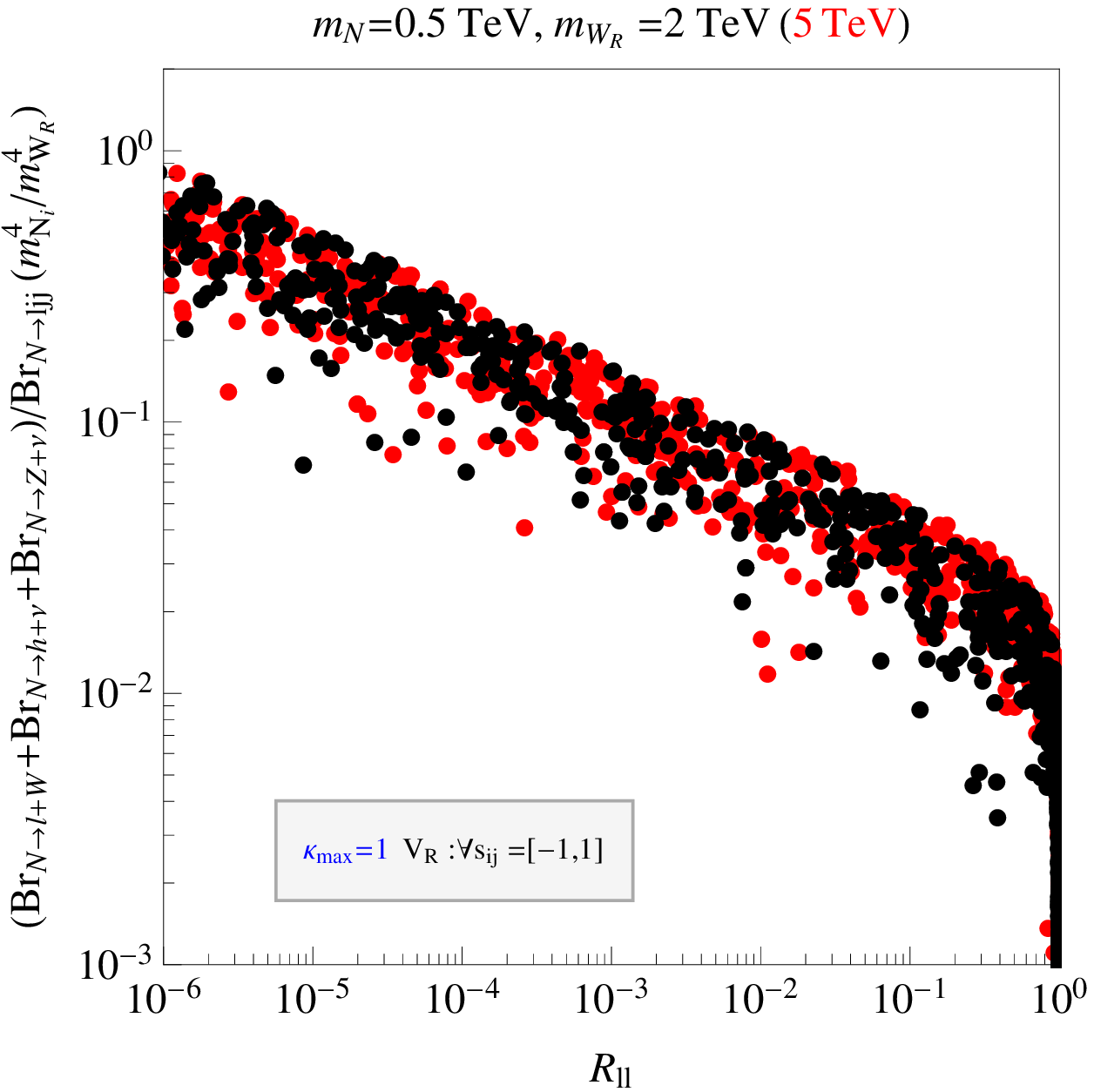}
\caption{\label{fig:RatvRll} Left panel: the SS to OS ratio $R_{ll}$
  versus $\Delta M/\Gamma$. Right panel: the sum of the branching
  ratios of $N$ decays to SM bosons divided by the branching ratio to
  $ljj$, versus $R_{ll}$.  The numerator has been rescaled by
  $(m_{N}/m_{W_R})^4$ to compensate for the $W_R$-propagator
  suppression in the denominator. Black points are for $m_{W_R}=2$
  TeV, red points for $m_{W_R}=5$ TeV, and $m_{N}=0.5\,$TeV. }
\end{figure}

Up to now we have kept the values of the entries of the $V_R$ and
${\cal R}$ matrices in the parametrization given in \eq{eq:Ddef} fixed
at some arbitrary constant values.  We recall that $V_R$ is a unitary
matrix with three angles and three phases, while ${\cal R}$ is complex
orthogonal and can be defined in the usual way in terms of 
$\sin$ and $\cos$ of three complex angles
$\zeta_i$. For our numerical scan, we parametrize these 
angles as: 
\begin{equation}\label{eq:defzi}
\zeta_i = \kappa \cdot e^{2i\pi\, x_i} ,
\end{equation}
with $x_i$ a randomly generated real number $\in [0,1]$, and
$\kappa \in [0,\kappa_{\rm max}]$.  The upper limit $\kappa_{\rm max}$
represents a measure of how much fine tuning is allowed in the
parametrization \eq{eq:Dsing} in order to allow for particularly large
values of $m_D$ (or alternatively of the Yukawa couplings generating $m_D$)
while still respecting all the constraints from low energy neutrino
data. For $\kappa_{\rm max} \lsim 1$ there is no fine tuning: all the
tree-level formulas presented above remain valid and in particular
loop corrections to neutrino masses and mixing angles remain at the
level of few percent. However, for $\kappa_{\rm max} \gsim 2-3$, 
similarly large values of $\kappa$ become possible and the corresponding
results 
would be
highly questionable, since the tree-level approximation starts to
break down and in particular, when loop corrections are taken into
account, some low energy neutrino parameters might well drop out the
experimentally allowed range.  We have then plotted the results in
fig.~(\ref{fig:RatvRll}) adopting the educated choice
$\kappa_{\rm max} =1$.

In the left panel of fig.~\ref{fig:RatvRll} we depict  $R_{ll}$ versus
$\Delta M/\Gamma$ for some arbitrary value of the heavy neutrino mass,
scanning randomly over the entries in $V_R$ and ${\cal R}$.  We see
that for $\Delta M$ larger than a few times $\Gamma$, $R_{ll}$
approaches rapidly the Majorana limit $R_{ll}=1$. This result is
independent of the absolute mass scale of the heavy neutrinos.

As we have already noticed, the expected widths for the heavy neutrino
decays are too small to be directly measured at the LHC (see
fig.~\ref{fig:Gami}).  However, the ratio of two-body versus
three-body $N$ decays can be measurable.  At fixed values of $m_N$ and
$m_{W_R}$ this ratio is controlled by the value of $\mu$, which also
fixes the mass splitting of the quasi-Dirac neutrino pair, therefore
we can expect a correlation between the ratio of two body versus three
body $N$ decays, and $R_{ll}$. This is shown in the right panel in
fig.~(\ref{fig:RatvRll}) where this ratio is plotted versus $R_{ll}$
(summed over lepton flavors).  The sum of the two body decays in the
numerator of the ratio ($y$-axis in the right panel), has been
rescaled by $(m_{N}/m_{W_R})^4$ to compensate for the $W_R$-propagator
suppression for the three body decay. This renders the correlation
between the two observables nearly independent of the values of the
$W_R$ and $N$ masses.  As the figure shows, if a large value
$R_{ll} \sim 1$ is measured, the present scenario predicts that the
rate for decays into SM bosons should be smaller than a few percent of
the rate for three body decays times $(m_{N}/m_{W_R})^4$.  On the
other hand, if a small value $R_{ll} \lsim 10^{-2}$ is measured (or an
upper bound of the same order is set), the prediction is that a
sizeable fraction of RH neutrino decays should proceed via on-shell SM
bosons.  
Note that this correlation does not depend on the type of light
neutrino spectrum (normal versus inverted hierarchy).
Thus, the inverse seesaw not only allows for generic values
$R_{ll} < 1$, but it also implies a testable correlation between
$R_{ll}$ and the RH neutrino decay modes.

As we have already said, the results depicted in the plots have been
obtained by summing over the the final state lepton flavors.  However,
given that the mixing matrices controlling the flavor composition of
the dilepton final states are in principle generic, different flavor
final states such as $\mu e jj$ can naturally occur with large
branching ratios, while respecting the full set of low energy
constraints (we have checked that numerically generated dilepton
samples do not show suppressions of different flavor dilepton events).
Thus, we stress again that SS and OS dilepton events of different
flavors should be included as a potential contribution to the signal
and, most importantly, they should not be used as an estimate of the
backgrounds in experimental analyses.  In the attempt of scrutinizing
further lepton flavor violating (LFV) effects in the inverse seesaw
scenario, we have also calculated branching ratios for low energy LFV
processes, the most relevant of which is Br($\mu\to e \gamma$).  We
have found that Br($\mu\to e \gamma$) can provide additional relevant
constraints only for very small values of $\mu$ ($\mu \ll 10^{-6}$
GeV), which corresponds to the regime in which the pure Dirac limit is
approached and $R_{ll}\approx 0$ is expected.

All in all, the main conclusion of this section is that LR models
equipped with an inverse seesaw mechanism for the light neutrino
masses naturally yield pairs of quasi-Dirac RH neutrinos. In the
specific region of parameter space corresponding to
$\Delta M \approx \Gamma$, the ratio $R_{ll}$ can have any value
within the range [0,1]. Moreover, this value correlates in a specific
way with the value of the ratio between two-body and three-body RH
neutrino decays, and gross violations of this prediction would disfavor
the scenario, and possibly rule it out.

\section{Summary}
\label{sect:sum}

In this paper we have discussed signals of LNV that could originate in
scenarios with quasi-Dirac neutrinos, that can be defined as a pair of
Majorana neutrinos for which a mass splitting much smaller than their
average mass is induced by small LNV terms.  In particular, we have
focused on the ratio of same-sign to opposite-sign dilepton events
$R_{ll}$, which is the most promising LNV observable for experimental
searches at the LHC. It is well known that if the dilepton events
originate from production/decays of heavy Majorana neutrinos, then
$R_{ll}=1$ is expected. We have shown that in the quasi-Dirac case, in
the regime in which the mass splitting $\Delta M$ between the pair of
heavy RH neutrino resonances becomes of the order of their widths, 
 any value within the interval $R_{ll}\in [0,1]$ is possible,
and $R_{ll}=0$ is approached in the limit $\Delta M/\Gamma \to 0$ which
defines the pure Dirac limit of the quasi-Dirac neutrino pair.  It
is then clear that an experimental result $R_{ll} < 1 (\neq 0)$
could provide valuable information about the mechanism of generation
of the light neutrino masses.

We stress that our main result on $R_{ll}$ does not depend on the
particular model realization of the quasi-Dirac neutrino scenario
(other features, as for example the total event rate for heavy
neutrino production, obviously do depend on the specific model). 
For definiteness we have carried out our discussion in the framework
of a LR symmetric model equipped with an inverse seesaw mechanism,
since this setup appears to be of prominent experimental interest in
view of the ongoing searches for signals of LNV and of RH neutrinos at
the LHC.  In discussing the LHC phenomenology, we have pointed out
that specific values of $R_{ll}\neq 0,1$ can be correlated with
special features of observables in the decay modes of the heavy
neutrinos, and this correlation can help to test the scenario.  Last
but not least, in developing our analysis we have introduced a new
parametrization of the inverse seesaw which allows to scan the
parameter space of the fundamental theory while automatically
respecting all the phenomenological constraints of the low energy
effective theory. The use of this parametrization has proven to be
very convenient in carrying out our numerical study.

\section*{Acknowledgments}
We thank R.N. Mohapatra and P. S. Bhupal Dev for discussion about their  
results.  The work of E.N. is supported in part from the
research grant ``Theoretical Astroparticle Physics" number 2012CPPYP7
under the program PRIN 2012 funded by the Italian ``Ministero dell'
Istruzione, Universit\`a e della Ricerca'' (MIUR) and from the INFN
``Iniziativa Specifica'' Theoretical Astroparticle Physics (TAsP-LNF).
This work was supported by the Spanish grants FPA2014-58183-P,
Multidark CSD2009-00064 and SEV-2014-0398 (from the \textit{Ministerio
  de Economía y Competitividad}), as well as PROMETEOII/2014/084 (from
the \textit{Generalitat Valenciana}).

\end{document}